\let\useblackboard=\iftrue
%
%
\newfam\black
\input harvmac.tex

\lref\topchange{P. Aspinwall, B. Greene, and D. Morrison, 
``Calabi-Yau Moduli Space, Mirror Manifolds and Spacetime Topology Change in
     String Theory'', Nucl.Phys. B416 (1994) 414-480, hep-th/9309097.}
\lref\edphases{E. Witten, ``Phases of N=2 Theories in Two Dimensions'',
Nucl.Phys. B403 (1993) 159-222, hep-th/9301042.}
\lref\dk{J. Distler and S. Kachru, ``(0,2) Landau-Ginsburg Theory'',
Nucl. Phys. B413 (1994) 213, hep-th/9309110.}
\lref\gh{O. Ganor and A. Hanany, 
``Small $E_8$ Instantons and Tensionless Non-critical Strings'',
Nucl.Phys. B474 (1996) 122-140, hep-th/9602120.}
\lref\swsix{N. Seiberg and E. Witten, ``Comment on
String Dynamics in Six Dimensions'', 
Nucl.Phys. B471 (1996) 121-134, hep-th/9603003.}
\lref\horwit{P. Horava and E. Witten, 
``Heterotic and Type I String Dynamics from Eleven Dimensions'',
Nucl.Phys. B460 (1996) 506-524, hep-th/9510209.}
\lref\silvwit{E. Silverstein and E. Witten, ``Criteria for Conformal
Invariance of (0,2) Models'', B444 (1995) 161, hep-th/9503212.}
\lref\gsw{M. Green, J. Schwarz, and E. Witten,  {\it Superstring Theory},
v. 2, Cambridge University Press (1987).}
\lref\seireview{K. Intriligator and N. Seiberg, ``Lectures on 
supersymmetric gauge theories and electric-magnetic duality'',
Nucl.Phys.Proc.Suppl. 45BC (1996) 1-28, hep-th/9509066.}
\lref\sw{N. Seiberg and E. Witten,
``Electric-Magnetic Duality, Monopole Condensation, and
Confinement in N=2 Supersymmetric Yang-Mills Theory'', 
Nucl.Phys. B426 (1994) 19-52, hep-th/9407087.}
\lref\Nfour{see e.g. 
E. Witten, ``String Theory Dynamics in Various Dimensions'',
Nucl.Phys. B443 (1995) 85-126, hep-th/9503124.}
\lref\Ntwo{see e.g. P. Green and T. Hubsch, ``Phase Transitions among
(Many of) Calabi-Yau Compactifications'', Phys. Rev. Lett. 61 (1988) 1163;
P. Candelas, P. Green, and T. Hubsch, ``Finite Distances between 
Distinct Calabi-Yau Vacua'', Phys. Rev. Lett. 62 (1989) 1956;
A. Strominger, ``Massless Black Holes and Conifolds in String Theory'',
Nucl. Phys. B451 (1995) 96, hep-th/9504090; 
B. Green, D. Morrison, and A. Strominger, ``Black Hole Condensation
and the Unification of String Vacua'', Nucl. Phys. B451 (1995) 109,
hep-th/9504145.}
\lref\None{see e.g. S. Kachru, N. Seiberg, and E. Silverstein,
``SUSY Gauge Dynamics and Singularities of 4d N=1 String Vacua'',
Nucl.Phys. B480 (1996) 170-184, hep-th/9605036;
Matthias Klein and Jan Louis, ``Singularities in d=4, N=1 
Heterotic String Vacua'', hep-th/9707047.}
\lref\bfss{T. Banks, W. Fischler, S. Shenker, and L. Susskind,
``M Theory As A Matrix Model: A Conjecture'', 
Phys.Rev. D55 (1997) 5112-5128,  hep-th/9610043.}
\lref\kschir{S. Kachru and E. Silverstein, 
``Chirality Changing Phase Transitions in 4d String Vacua'',
hep-th/9704185.}
\lref\threed{O. Aharony, A. Hanany, K. Intriligator, N. Seiberg, M.J.
     Strassler, ``Aspects of N=2 Supersymmetric Gauge 
Theories in Three Dimensions'', 
Nucl.Phys. B499 (1997) 67-99, hep-th/9703110.}
\lref\aldezable{G. Aldazabal, A. Font, L.E. Ibanez, A.M.Uranga, G. Violero,
``Non-Perturbative Heterotic D=6,4 Orbifold Vacua'', 
hep-th/9706158.}
\lref\fmw{Robert Friedman, John Morgan, and Edward Witten,
``Vector Bundles And F Theory'', Commun.Math.Phys. 187 (1997) 679-743,
hep-th/9701162.}
\lref\bershad{Michael Bershadsky, Andrei Johansen, Tony Pantev and
Vladimir Sadov, ``On Four-Dimensional Compactifications of F-Theory'',
hep-th/9701165.}

\noblackbox

\def\Title#1#2{\rightline{#1}
\ifx\answ\bigans\nopagenumbers\pageno0\vskip1in%
\baselineskip 15pt plus 1pt minus 1pt
\else
\def\listrefs{\footatend\vskip 1in\immediate\closeout\rfile\writestoppt
\baselineskip=14pt\centerline{{\bf References}}\bigskip{\frenchspacing%
\parindent=20pt\escapechar=` \input
refs.tmp\vfill\eject}\nonfrenchspacing}
\pageno1\vskip.8in\fi \centerline{\titlefont #2}\vskip .5in}

\ifx\answ\bigans\def\tcbreak#1{}\else\def\tcbreak#1{\cr&{#1}}\fi
\useblackboard
\message{If you do not have msbm (blackboard bold) fonts,}
\message{change the option at the top of the tex file.}
\font\blackboard=msbm10 scaled \magstep1
\font\blackboards=msbm7
\font\blackboardss=msbm5
\textfont\black=\blackboard
\scriptfont\black=\blackboards
\scriptscriptfont\black=\blackboardss

\else

\fi
\def\yboxit#1#2{\vbox{\hrule height #1 \hbox{\vrule width #1
\vbox{#2}\vrule width #1 }\hrule height #1 }}
\def\fillbox#1{\hbox to #1{\vbox to #1{\vfil}\hfil}}
\def\ybox{{\lower 1.3pt \yboxit{0.4pt}{\fillbox{8pt}}\hskip-0.2pt}}

\def\comments#1{}

\def\II{\relax{I\kern-.07em I}}

\def\IZ{\relax\ifmmode\mathchoice
{\hbox{\cmss Z\kern-.4em Z}}{\hbox{\cmss Z\kern-.4em Z}}
{\lower.9pt\hbox{\cmsss Z\kern-.4em Z}}
{\lower1.2pt\hbox{\cmsss Z\kern-.4em Z}}\else{\cmss Z\kern-.4em
Z}\fi}
\def\IB{\relax{\rm I\kern-.18em B}}
\def\IC{{\relax\hbox{$\inbar\kern-.3em{\rm C}$}}}
\def\ID{\relax{\rm I\kern-.18em D}}
\def\IE{\relax{\rm I\kern-.18em E}}
\def\IF{\relax{\rm I\kern-.18em F}}
\def\IG{\relax\hbox{$\inbar\kern-.3em{\rm G}$}}
\def\IGa{\relax\hbox{${\rm I}\kern-.18em\Gamma$}}
\def\IH{\relax{\rm I\kern-.18em H}}
\def\II{\relax{\rm I\kern-.18em I}}
\def\IK{\relax{\rm I\kern-.18em K}}
\def\IP{\relax{\rm I\kern-.18em P}}

\font\cmss=cmss10 \font\cmsss=cmss10 at 7pt
\def\IR{\relax{\rm I\kern-.18em R}}

\def\IZ{\relax\ifmmode\mathchoice
{\hbox{\cmss Z\kern-.4em Z}}{\hbox{\cmss Z\kern-.4em Z}}
{\lower.9pt\hbox{\cmsss Z\kern-.4em Z}}
{\lower1.2pt\hbox{\cmsss Z\kern-.4em Z}}\else{\cmss Z\kern-.4em
Z}\fi}
\def\IB{\relax{\rm I\kern-.18em B}}
\def\IC{{\relax\hbox{$\inbar\kern-.3em{\rm C}$}}}
\def\ID{\relax{\rm I\kern-.18em D}}
\def\IE{\relax{\rm I\kern-.18em E}}
\def\IF{\relax{\rm I\kern-.18em F}}
\def\IG{\relax\hbox{$\inbar\kern-.3em{\rm G}$}}
\def\IGa{\relax\hbox{${\rm I}\kern-.18em\Gamma$}}
\def\IH{\relax{\rm I\kern-.18em H}}
\def\II{\relax{\rm I\kern-.18em I}}
\def\IK{\relax{\rm I\kern-.18em K}}
\def\IP{\relax{\rm I\kern-.18em P}}

\font\cmss=cmss10 \font\cmsss=cmss10 at 7pt
\def\IR{\relax{\rm I\kern-.18em R}}

\def\tilde{\widetilde}
\def\frac#1#2{{{#1} \over {#2}}}

\Title{\vbox{\baselineskip12pt\hbox{hep-th/9709209}
\hbox{SLAC-PUB-7710}}}
{\vbox{\centerline{Closing the Generation Gap}
\centerline{}
\centerline{}}}

\centerline{Eva Silverstein}

\smallskip
\smallskip
\smallskip
\centerline{evas@slac.stanford.edu}
\centerline{Stanford Linear Accelerator Center}
\centerline{Stanford University}
\centerline{Stanford, CA 94309, USA}
\bigskip
\bigskip
\noindent

I describe recent examples of phase transitions in
four-dimensional M theory vacua in which
the net generation number changes.  
There are naive obstructions to transitions lifting chiral
matter, but loopholes exist which enable us to avoid them.
I first review how chirality arises in the heterotic
limit of M theory, previously known forms
of topology change in string theory, and chirality-changing
phase transitions in six dimensions.  This leads to the construction of
the four-dimensional examples, which involve wrapped M-theory
fivebranes at an $E_8$ wall.
(Talk presented at Strings '97, Amsterdam.)

\Date{September 1997}

Chiral fermions play a large role in low-energy particle physics.
The fermion mass term in a Lagrangian is given by
\eqn\Lmass{{\cal L}_m=-m(\overline\psi_L\psi_R+\overline\psi_R\psi_L).}
This ensures that fermions
in complex representations of the gauge group (as in the Standard
Model) do not have 
gauge and Lorentz invariant mass terms, as long as the gauge
group which distinguishes the left and right-handed fermions
remains unbroken.

In string theory, or more generally M theory, chirality is related
to the topology of the space on which the strings propagate.
For example in the limit of weakly coupled
$E_8\times E_8$ Heterotic strings,
chiral matter in four dimensions is obtained in the following way \gsw.
To obtain a four-dimensional vacuum, we take the strings to
propagate on a spacetime of the form
\eqn\hetcomp{M_4\times(X_6,V)}
where $M_4$ is four-dimensional Minkowski space, and $(X_6,V)$ is
a manifold $X_6$ with a vector bundle $V$ satisfying the string
equations of motion.  For example let us take the components
$A_i$ of the ten-dimensional gauge bosons along $X_6$ to have
vacuum expectation values in $SU(3)$.  Then the adjoint
of $E_8$ (the representation under which the ten-dimensional
gauge fields transform) decomposes under the surviving unbroken
gauge group $E_6$ times the broken $SU(3)$ as
\eqn\decomp{{\bf 248}\to {\bf (1,8)+(78,1)+(27,3)+(\overline{27},
\overline{3})}.}

So the net number of generations of $E_6$ is
\eqn\ngen{N_{gen}=n_{{\bf 27}}^R-n_{{\bf 27}}^L 
= n_{{\bf 3}}^+-n_{{\bf 3}}^-
=Index_{\bf 3}({\cal D}_{X_6}) .}
where ${\cal D}_{X_6}$ is the gauge-covariant Dirac operator on $X_6$.  
In other words, the chirality in the four-dimensional Minkowski
space is related to the chirality on the compactification manifold.
Now $N_{gen}$ is a topological invariant of $(X_6,V)$.  Therefore
it cannot change under smooth deformations of $(X_6,V)$.  

This leads to two apparent (related) obstructions to unifying 
M-theory vacua, which in general have different net
generation numbers:
(i) We have seen that in order to change
$N_{gen}$ the compactification must become singular, since
the Dirac index must change. More seriously perhaps,
(ii) From \Lmass\ we saw that chirality change cannot happen in
ordinary low-energy field theory (i.e. weakly coupled Lagrangian
field theory), where chiral matter cannot get lifted.  

Recent progress has uncovered 
sensible physical resolutions to singularities in compactifications,
so (i) is not a problem in and of itself.  Indeed, we now
know how to interpret many examples of singular geometries (i.e. singular
solutions to classical general relativity).  
There are examples \topchange\edphases\dk\ of topology change at
the classical level in string theory.  This is possible because
point particle classical spacetime geometry is at best approximately
valid in the large-radius limit.  One can achieve topology 
change by going through regions in the moduli space of the 
$2d$ CFT defining the string vacuum where $\alpha^\prime$ corrections
and worldsheet instanton effects are large so classical geometry
breaks down. The singularity can be avoided in going between
phases of the compactification with different topology at large radius.

More similar to our situation are cases where in order to effect
topology change one must grapple with a singularity which remains
after all classical stringy corrections are included.
The lesson has been that physics is nonsingular as long as all light
degrees of freedom associated with the singularity are taken into account.
This has been a central principle in analysis of supersymmetric
gauge theory dynamics \seireview\sw.  The phenomenon has been similarly
observed in string theory compactifications with $N=4$ \Nfour, 
$N=2$ \Ntwo, and $N=1$ \None\ supersymmetry in four dimensions.  
All of these examples of singularity resolution
involved conventional low-energy quantum
field theories realized in various string compactifications.

As for problem (ii), the impossibility of lifting chiral matter
in Lagrangian field theory, the loophole is to consider regions
of moduli space in which the low-energy effective quantum field
theory is not weakly coupled.  Many nontrivial interacting
fixed point quantum field theories have been found in the moduli
space of various compactifications of M-theory.  We find
\kschir\ examples in which the net number of generations changes
upon going through a locus in the moduli space where the effective
theory is a nontrivial RG fixed point. 

\newsec{Review of chirality change in six dimensions}

Indeed, something similar occurs in six dimensional (1,0) supersymmetric
theories \gh\swsix\ obtained by considering M theory compactified
on $S^1/Z_2\times K3$.  At the end of the interval $S^1/Z_2$ live
ten-dimensional $E_8$ gauge bosons \horwit.  In general their
components along $K3$ comprise the connection of a holomorphic vector bundle
$\tilde V$ over $K3$, in other words a configuration of instantons living
on $K3$.  In order to have a perturbative heterotic description, we
need 24 instantons on $K3$ at the end of the interval.  Let us
take all the instantons to lie in an $SU(2)$ subgroup of $E_8$, leaving
an unbroken $E_7$ gauge group in spacetime.  

The spectrum consists of 20 ${1\over 2}{\bf 56}$'s of $E_7$ and
65 singlet moduli (45 moduli of the instanton bundle and 20 moduli
of $K3$).  Each instanton has a scale size modulus.  Let us consider
shrinking a collection of instantons to zero size.  As explained
in \gh\swsix\ the shrunk instantons correspond to M-theory fivebranes
at the end of the interval.  There is then another phase in which
the M-theory fivebranes move off into the eleven-dimensional bulk.

The spectrum in this second phase includes 
for each fivebrane a (1,0) tensor multiplet,
whose real scalar parameterizes the distance of the fivebrane
from the end of the interval, as well as a hypermultiplet whose
four scalars give its position on $K3$.
In order to shrink an instanton and move into this second phase,
one loses the other collective coordinates of the Yang-Mills instanton:
one hypermultiplet containing the scale size and one ${1\over 2}{\bf 56}$
of $E_7$.  
This is in accord with the anomaly condition which requires the addition of
one tensor to be compensated by losing 29 hypermultiplets.
As explained in \swsix, this transition changing the number of
tensor multiplets cannot happen in weakly coupled
Lagrangian field theory.
For the four-dimensional application we are interested in here,
the main result we will need from six dimensions is the loss
of charged matter (one ${1\over 2}{\bf 56}$ per instanton).

\newsec{Changing $N_{gen}$}

We can return to four-dimensional physics by considering M-theory
on $S^1/Z_2\times (X_6,V)$ where $X_6$ is taken to be a $K3$ fibration,
and $V$ as above is taken to be an $SU(3)$ bundle.   
In other words, $(X_6, V)$ has the structure of a family
of $(K3,\tilde V)$'s varying over a ${\bf CP}^1$ base.
The four-dimensional theory is a sort of ``twisted'' dimensional
reduction of the six dimensional theory studied in \S2\ down
to four dimensions.  The massless spectrum is obtained by
finding zero eigenstates of the Dirac operator on 
the ${\bf CP}^1$ base, where the Dirac operator includes 
contributions arising from the variation of the fiber $(K3,\tilde V)$
over the base.

In \None\kschir\ we studied a set of $K3$ fibrations using the
linear sigma model approach \edphases\ for the heterotic compactifications.
In particular, in \kschir\ we found the following pattern in
the four dimensional spectrum.  Each ${1\over 2}{\bf 56}$ in the
six dimensional theory descends to chiral matter (either 
${\bf 27}$'s of $E_6$ or a ${\bf \overline{27}}$ of $E_6$).
For more details on the analysis of the spectrum, see
the paper \kschir.
In addition, there is charged matter that does not descend from
the six dimensional theory, in other words matter that is associated
to the singular fibers.  

Now we can shrink one of the instantons in the generic fiber
theory $(K3,\tilde V)$, and move it off the end of the interval
as an M-theory fivebrane wrapped on the base ${\bf CP}^1$.
Recall from \S2\ that in the corresponding six dimensional theory
we lost a ${1\over 2}{\bf 56}$ in the transition.  This means
that in four dimensions we lose the {\it chiral} matter that
descends from the ${1\over 2}{\bf 56}$.

At the origin between the two branches there is, as in 
six dimensions, a nontrivial interacting fixed point theory.
In six dimensions, the presence of stringlike BPS states coming
down to zero tension at the origin signals the presence of
a CFT there.  
This remains true after our fibration down to four dimensions:
in particular, the base ${\bf CP}^1$ does not admit string winding
modes so the unwrapped strings remain the lightest degrees of freedom
as we approach the origin.

One can identify the charged matter that gets removed in
the transition using the linear sigma model description of
singularities \edphases.  At the singularity in
the vector bundle that we studied above, the
linear sigma model target space becomes noncompact, developing
a long tube.  The vertex operators can be identified
\silvwit, and those which are supported down
this tube are chiral (i.e. either ${\bf 27}$'s or ${\bf \overline{27}}$'s,
but not both).  The throat carries the information about
the singularity and its nonperturbative resolution by
a nontrivial interacting conformal field theory.  So only the
states which are supported down the throat are involved in the transition.

The singular fibers also do not appear to change this result,
for the following reasons.  Generically, the singularities in
the singular fibers are at points on the $K3$.  These points
are generically separated from the singularity induced by
shrinking the instanton.  Furthermore an F-theory analysis 
of the $6d$ theory shows that removing the small instanton 
from the end of the interval does not introduce additional singularities.  

So the conclusion is that the
net number of generations $N_{gen}$ changes in the transition!

\newsec{Instanton Effects}

So far we have been doing essentially a Kaluza-Klein analysis,
i.e. looking at scales below $1/R$ where $R$ is the radius
of the base ${\bf CP}^1$, but ignoring instanton effects.
There are ordinary heterotic string instantons arising from
fundamental heterotic string worldsheets wrapping the base
${\bf CP}^1$ $B$.  In eleven dimensions, this corresponds to the
membrane worldvolume stretching between the ends of the interval 
and wrapping $B$.  With the fivebrane in the middle of the
interval, we have two additional types of strings:  those 
arising from the membrane stretched between the fivebrane
and either end of the interval.  

The fundamental string instanton effects go like 
$e^{-{R^2\over\alpha^\prime}}$.  The others go like 
$e^{-R^2\phi}$ and $e^{-{R^2\over\alpha^\prime}+R\phi}$,
where $\phi$ is the dimensional reduction of the
scalar in the $6d$ tensor multiplet.
Instanton contributions depend on zero-mode counting, which
depends on how $V$ restricts to $B$.  There are cases for
which instanton effects to not contribute to the superpotential
for singlet moduli.  In these cases the phase transition proceeds
at zero energy as described above.

There are other cases for which there are nontrivial instanton 
effects.  These can be understood by considering a T-dual
$SO(32)$ heterotic description in three dimensions
(after compactifying the above on a circle of radius $r$).  There
the small instanton singularity is nonperturbatively resolved
by a gauge symmetry enhancement.  Each small instanton carries
an $SU(2)$ gauge group with matter in the ${\bf (32,2)}$.  
The number of surviving doublets in four dimensions depends on how $V$ 
restricts to $B$ \None.  

In the case where two flavors survive the superpotential
was determined in \threed\ to be
\eqn\superthree{W=W_{tree}-YPfV+e^{-{R^2\over\alpha^\prime}}Y.}
where $Y=e^{R^2\phi}$ for large $R^2\phi$,
and $V_{ij}$ are gauge-invariant coordinates
on the moduli space.  The $3d$ and $4d$ gauge couplings are
related by ${1\over {rg_3^2}}={1\over g_4^2}={R^2\over\alpha^\prime}$.
We are interested in the limit $r\to 0$, $R^2\over\alpha^\prime$ fixed,
in order to return to the four-dimensional compactification of
the $E_8\times E_8$ heterotic string that we have been discussing.
As in \None, with an appropriate $W_{tree}$ one can reproduce
the pole in the Yukawa couplings which occurs for these compactifications.  

In this case there is a term $|Pf V-e^{-{R^2\over\alpha^\prime}}|^2$
in the potential energy.  In order to shrink an instanton (which
here corresponds to taking $V\to 0$) there is a cost in energy
proportional to 
${1\over{\sqrt{\alpha^\prime}}}e^{-2{R^2\over\alpha^\prime}}$.
So in this case the transition can occur, but only by going
over a small energy barrier.

\newsec{Conclusions}

We have explained how the naive obstructions to unifying vacua
with different net generation numbers can be overcome, and
we gave a class of examples of chirality-changing phase transitions
in four dimensions (see also the recent examples of \aldezable\
in orbifold models).
The phenomenon exhibited here presumably occurs quite generally; it
would be very interesting to understand whether in fact one
can connect all vacua with $N\le 1$ supersymmetry at low energies.
For this an F-theory analysis could be quite instructive,
once that approach is developed more fully for four-folds
(see e.g. \fmw\bershad ).

Perhaps the next ``in principle'' question to ask along these lines is whether
there could be any physical process which changes the 
number of supersymmetries.  At low energies, any such transition
would necessarily involve gravity, since the number of gravitinos
would have to change in the transition.  The moduli space of
theories with $N\ge 4$ supersymmetry is so constrained that
there does not seem to be any room for such a phase transition
at low energies (the singularities in these moduli spaces are
all accounted for in weakly coupled Lagrangian field theory at
low energies).  This pushes the question to high energies,
where we probably need a background-independent formulation
of M-theory to really address it; perhaps the Matrix theory \bfss\ can
provide some clues.  

\bigskip
{\bf Acknowledgements}

This talk was based on a paper \kschir\ with S. Kachru.  I am 
also indebted to O. Aharony, P. Aspinwall, T. Banks, P. Candelas,
J. Distler, M. Douglas, O. Ganor, A. Hanany, J. Louis, N. Seiberg,
S. Shenker, and E. Witten for discussions on this and related topics.
I would also like to thank the organizers of Strings '97 for a very
stimulating conference.

\listrefs
\end